\documentclass[12pt]{article}
\usepackage{amssymb,amsmath}
\begin{document}
%%%%%%%%%%%%%%%%%%%%%%%%%%%%%%%%%%%%%%%%%%%%%%%%%%%%%%%%%%%%%%%%%%%%%
%%%%%%%%%%%%%%      TITLE PAGE     %%%%%%%%%%%%%%%%%%%%%%%%%%%%%%%%%%
%%%%%%%%%%%%%%%%%%%%%%%%%%%%%%%%%%%%%%%%%%%%%%%%%%%%%%%%%%%%%%%%%%%%%

\sloppy
\title
%{\hfill{\normalsize\sf FIAN/TD/01-15}    \\
 %           \vspace{1cm}
{\Large  On summation of Kapteyn series }

\author
 {
       A.I.Nikishov
          \thanks
             {E-mail: nikishov@lpi.ru}
  \\
               {\small \phantom{uuu}}
  \\
           {\it {\small} I.E.Tamm Department of Theoretical Physics,}
  \\
               {\it {\small} P.N.Lebedev Physical Institute, Moscow, Russia}
  \\
  %       {\it {\small} 117924, Leninsky Prospect 53, Moscow, Russia.}
 }
%
%--------------------------------------------------------------------
\maketitle
\begin{abstract}
We consider the summation of Kapteyn series of several kinds and
obtain some relations among the sums. We paid special attention to
cases when sums are transcendental functions. Their asymptotic
behaviors are obtained. In some cases the integral representations
for sums are found. As an application we find the radiation free
lifetime of an electron moving in a constant external field. It is
noted that the laboratory lifetime increases with energy for
ultrarelativistic electron in ultrastrong external field.
\end{abstract}

\section{Introduction}
Kapteyn series linear in Bessel function we call linear series.
Series, sum of which is given by an algebraic function, we call
algebraic. Otherwise we call it transcendental. Starting with some
linear algebraic series we obtain their sums. Then we consider
transcendental linear series and find integral representation for
some of them. We investigate the asymptotic behavior of these
transcendental functions. Then we do the same with bilinear
transcendental functions. We establish some relations among the
sums. Finally, as an example of application we get the radiation
free lifetime of an electron moving in a constant external field. We
assume that $z^2$ (or $x^2$) are less the unity.
\section{Linear algebraic series}
We start with linear series because some of bilinear series can be

obtained from linear one with the help of the relation
$$
J^2_n(nx)=\frac{2}{\pi}\int^{\pi/2}_{0}d\varphi
J_{2n}(2nx\cos\varphi),                                \eqno(2.01)
$$
see (6.681.9) in Gradshteyn and Ryzhic, 2000.

Most of the  the results we are interested in this section  are contained
in Watson, 1966, see also Lerche and Tautz, 2007. We want to
stress here that essentially all can be obtained from three
relations:
$$
\left(z\frac{d}{dz}\right)^2\sum^{\infty}_{m=1}\frac{a_m}{m^2}J_m(mz)=(1-z^2)\sum^{\infty}_{m=1}a_mJ_m(mz),
                                                             \eqno (2.02)
$$

$$
\sum^{\infty}_{m=1}J_m(mz)=\frac12\left(\frac{1}{1-z}-1\right)=\frac{z}{2(1-z)},
                                                       \eqno(2.03)
$$

$$
\sum^{\infty}_{m=1}(-1)^mJ_m(mz)=\frac12\left(\frac{1}{1+z}-1\right).
                                                      \eqno(2.04)
$$
The sum (difference) of (2.03) and (2.04) gives sum over even $m=2n$
(odd $m=2n+1$):
$$
\sum^{\infty}_{n=1}J_{2n}(2nz)=\frac{z^2}{2(1-z^2)}, \eqno(2.05)
$$

$$
\sum^{\infty}_{n=1}J_{2n+1}((2n+1)z)=\frac{z}{2(1-z^2)},\eqno(2.06)
$$

Another relations are obtained by differentiation:
$$
\sum^{\infty}_{n=1}2nJ'_{2n}(2nz)=\frac{z}{(1-z^2)^2}, \eqno(2.07)
$$

$$
\sum^{\infty}_{n=1}(2n+1)J'_{2n+1}((2n+1)z)=\frac{1+z^2}{2(1-z^2)^2},\eqno(2.08)
$$
Differentiating (2.07) we get
$$
\sum^{\infty}_{n=1}(2n)^2J"_{2n}(2nz)=\frac{1+3z^2}{(1-z^2)^3},
                                               \eqno(2.09)
$$
Using here the Bessel equation
$$
2nJ''_{2n}(2nz)=\left(\frac{1}{z^2}-1\right)2nJ_{2n}(2nz)-
\frac1zJ'_{2n}(2nz)
                                                          \eqno(2.10)
$$
we find
$$
\left(\frac{1}{z^2}-1\right)\sum^{\infty}_{n=1}(2n)^2J_{2n}(2nz)-
\frac1z\sum^{\infty}_{n=1}2nJ'_{2n}(2nz)=\frac{1+3z^2}{(1-z^2)^3}.\eqno(2.11)
$$
We can write (2.10) in the form
$$
\left(\frac{d}{dz}+\frac1z\right)J'_{2n}(2nz)=\left(\frac{1}{z^2}-1\right)2nJ_{2n}(2nz).
\eqno(2.10')
$$
 Multiplying by $a_n$ and summing we get
$$
\left(\frac{d}{dz}+\frac1z\right)\sum^{\infty}_{n=1}a_nJ'_{2n}(2nz)=
\left(\frac{1}{z^2}-1\right)\sum^{\infty}_{n=1}a_n2nJ_{2n}(2nz).
\eqno(2.10'')
$$

 Using (2.03) in the r.h.s. of (2.02) with $a_m=1$ we get
$$
\left(z\frac{d}{dz}\right)^2\sum^{\infty}_{m=1}m^{-2}J_m(mz)=\frac
z2(1+z). \eqno(2.12)
$$
Integrating we find
$$
\sum^{\infty}_{m=1}m^{-2}J_m(mz)=\frac {z}{2}\left(1+\frac
{z}{4}\right).                                          \eqno(2.13)
$$
Similarly we get
$$
 \sum^{\infty}_{m=1}(-1)^mm^{-2}J_m(mz)=\frac {z}{2}\left(-1+\frac
{z}{4}\right).                                         \eqno(2.14)
$$
The sum and difference of (2.13) and (2.14) are given in Watson
book, see there eqs. 17.33(2) and 17.33(3). In similar manner we
find
$$
\sum^{\infty}_{n=1}\frac{1}{2n}J'_{2n}(2nz)=\frac z4,\quad
\sum^{\infty}_{n=1}\frac{1}{2n+1}J'_{2n+1}((2n+1)z)=\frac12.
                                                  \eqno(2.15)
$$
Putting $a_m=m^2$ in (2.02) we have
$$
\left(z\frac{d}{dz}\right)^2\sum^{\infty}_{m=1}J_m(mz)=
(1-z^2)\sum^{\infty}_{m=1}m^2J_m(mz).
                                                             \eqno (2.16)
$$
Using (2.03) in the l.h.s. and performing differentiation we get
$$
\sum^{\infty}_{m=1}m^2J_m(mz)=\frac{z}{2(1-z)^4}. \eqno(2.17)
$$
Similarly we obtain
$$
\sum^{\infty}_{m=1}(-1)^mm^2J_m(mz)=-\frac{z}{2(1+z)^4}. \eqno(2.18)
$$
Putting $a_m=m^4$ in (2.02) we have
$$
\left(z\frac{d}{dz}\right)^2\sum^{\infty}_{m=1}m^2J_m(mz)=
(1-z^2)\sum^{\infty}_{m=1}m^4J_m(mz).
                                                             \eqno (2.19)
$$
Using (2.17) in the l.h.s. and performing differentiation we get
$$
\sum^{\infty}_{m=1}m^4J_m(mz)=\frac{z(1+9z)}{2(1-z)^7}.
                                           \eqno(2.20)
$$
Putting $a_m=(-1)^m m^4$ in (2.02) we have
$$
\left(z\frac{d}{dz}\right)^2\sum^{\infty}_{m=1}(-1)^m m^2J_m(mz)=
(1-z^2)\sum^{\infty}_{m=1}(-1)^m m^4J_m(mz).
                                                             \eqno (2.21)
$$
Using (2.18) in the l.h.s. and  performing differentiation  we get
$$
\sum^{\infty}_{m=1}(-1)^m m^4J_m(mz)=-\frac{z(1-9z)}{2(1+z)^7}.
                                                       \eqno(2.22)
$$
Summing (2.20) and (2.22) we have
$$
\sum^{\infty}_{n=1}n^4J_{2n}(2nz)=\frac
{z^2(1+14z^2+21z^4+4z^6)}{2(1-z^2)^7},
                                           \eqno(2.23)
$$
see also Lerche and Tautz, 2007. The difference of (2.20) and
(2.22) gives
$$
\sum^{\infty}_{n=1}(2n+1)^4J_{2n+1}((2n+1)z)=\frac
{z(1+84z^2+350z^4+196z^6+9z^8)}{2(1-z^2)^7}.
                                           \eqno(2.24)
$$

\section{Linear transcendental series}

We start with the expression ($a\leq1$)
$$
\sum^{\infty}_{m=1}\frac{a^m}{m}J_m(mx)=
\frac{1}{2\pi}\int_{-\pi}^{\pi}d\theta
\sum^{\infty}_{m=1}\frac{1}{m}[ae^{-i(\theta-x\sin\theta)}]^m
$$
$$=
-\frac{1}{2\pi}\int_{-\pi}^{\pi}d\theta\ln[1-ae^{-i(\theta-x\sin\theta)}]\eqno(3.01)
$$
$$
=-\frac{1}{2\pi}\int_{0}^{\pi}d\theta\{\ln[1-ae^{-i(\theta-x\sin\theta)}]
+\ln[1-ae^{i(\theta-x\sin\theta)}] \}
$$
$$
=-\frac{1}{2\pi}\int_{0}^{\pi}d\theta\ln[1+a^2-2a\cos(\theta-x\sin\theta)].\eqno(3.02)
$$
We expand the integrand in (3.02) in powers of $x$:
$$
\Phi(b,e,y)\equiv\ln(b+e\cos(\theta+y))=\Phi(b,e,0)+
\left.\frac{\partial\Phi}{\partial y}\right|_{y=0}y+
\frac12\left.{\frac{\partial y^2\Phi}{\partial
y^2}}\right|_{y=0}y^2+\cdots,
$$
$$
b=1+a^2,\quad e=-2a,\quad y=-x\sin\theta.    \eqno(3.03)
$$
On the other hand we can calculate the l.h.s. of (3.01)
 straightforward:
$$
 \sum^{\infty}_{m=1}\frac{a^m}{m}J_m(mx)
 =\frac{ax}{2}+\frac{a^2x^2}{4}+(-a+3a^3)\frac{x^3}{2^4}+\cdots.
                                                 \eqno(3.04)
$$
Comparison of r.h.s. of (3.04) with (3.02) (together with (3.03))
gives integrals over $\theta$. Thus it should be
$$
\int_{0}^{\pi}d\theta \Phi(b,e,0)\equiv \int_{0}^{\pi}d\theta \ln(1+a^2-2a\cos\theta)=0.
\eqno(3.05)
$$
This is indeed so according to (4.397.16) in Gradsteyn and Ryzhic, 2000.

Similarly, it should be
$$
\frac{1}{2\pi}\int_{-\pi}^{\pi}d\theta
\left.\frac{\partial\Phi}{\partial
y}\right|_{y=0}x\sin\theta=\frac{ax}{2}.        \eqno(3.06)
$$

From (3.03) we have
$$
\frac{\partial\Phi}{\partial
y}=-\left.\frac{e\sin(\theta+y)}{b+e\cos(\theta+y)}\right|_{y=0}=
\frac{2a\sin\theta}{1+a^2-2a\sin\theta}.    \eqno(3.07)
$$
$$
\frac{\partial^2 \Phi}{\partial y^2}=-\left.\frac{e^2 +be
\cos(\theta+y)}{[b+e\cos(\theta+y)]^2}\right|_{y=0}= -\frac{e^2 +be
\cos\theta}{[b+e\cos\theta]^2}=
-\frac{e^2-b^2}{[b+e\cos\theta]^2}-\frac{b}{b+e\cos\theta}.
                                                    \eqno(3.08)
$$
So from (3.06) and (3.07) we have
$$
\frac{x}{2\pi}\int_0^{\pi}d\theta\frac{2a\sin^2\theta}{1+a^2-2a\sin\theta}
=\frac{ax}{2}. \eqno(3.09)
$$
This agrees with (2.5.16.26) in Prudnikov et al., 1986.

Now one more check that this method of obtaining integrals
over $\theta$ can be verified in simple cases. From (3.04), (3.02)
and (3.03) we expect that

$$
-\frac{1}{2\pi}\int_{0}^{\pi}d\theta\frac12\left.{\frac{\partial
y^2\Phi}{\partial y^2}}\right|_{y=0}\sin^2\theta
x^2=\frac{a^2x^2}{4}. \eqno (3.10)
$$
or, using (3.08),
$$
\frac{1}{2\pi}\int_{0}^{\pi}d\theta\frac12[\frac{e^2-b^2}{[b+e\cos\theta]^2}+
\frac{b}{b+e\cos\theta}]x^2\sin^2\theta=\frac{a^2x^2}{4}.   \eqno(3.11)
$$
The second term in square brackets is essentially (3.09):
$$
\int_0^{\pi}d\theta\frac{\sin^2\theta}{1+a^2-2a\sin\theta}
=\frac{\pi}{2}.                              \eqno(3.12)
$$
Differentiating this over $a$ we get
$$
\int_0^{\pi}d\theta\frac{(2a-2\cos\theta)\sin^2\theta}{[1+a^2-2a\cos\theta]^2}=0.
                                                \eqno (3.13)
$$
This may be written as
$$
\int_0^{\pi}d\theta\frac{\sin^2\theta}{[1+a^2-2a\cos\theta]^2}
\left(\frac{a^2-1}{a}+\frac{1}{a}(1+a^2-2a\cos\theta)
 \right)=0.                                         \eqno(3.14)
$$
From here
$$
\frac{1-a^2}{a}\int_0^{\pi}d\theta\frac{\sin^2\theta}{[1+a^2-2a\cos\theta]^2}=
\frac{1}{a}\int_0^{\pi}d\theta\frac{\sin^2\theta}{1+a^2-2a\cos\theta}.                                   \eqno(3.15)
$$
and using (3.12)
$$
\int_0^{\pi}d\theta\frac{\sin^2\theta}{[1+a^2-2a\cos\theta]^2}=\frac{\pi}{2(1-a^2)}. \eqno(3.16)
$$
Now, according to (3.03) we have
$$
e^2-b^2=-(1-a^2)^2, \quad b=1+a^2. \eqno(3.17)
$$
So the l.h.s. of (3.11) with the help of (3.12) and (3.16) takes the
form
$$
-\frac{x^2}{4\pi}[(1-a^2)-(1+a^2)]\frac{\pi}{2}=\frac{x^2a^2}{4}\eqno(3.18)
$$
with agreement with r.h.s. of (3.11).

Next we consider the special case when $a=1$ in (3.01), (3.02):
$$
\sum^{\infty}_{m=1}\frac{1}{m}J_m(mx)=-\frac{1}{2\pi}\int_0^{\pi}d\theta
\ln[2(1-\cos(\theta+y))]=-\frac{1}{\pi}\int_0^{\pi}d\theta
\ln[2\sin\left(\frac{\theta}{2}-\frac{x}{2}\sin\theta\right)].
                                                      \eqno(3.19)
$$
Instead of (3.04) we have
$$
\sum^{\infty}_{m=1}\frac{1}{m}J_m(mx)=\frac
x2+\frac{x^2}{2^2}+\frac{x^3}{2^3}+\frac{x^4}{2^2\cdot3}+\frac{23x^5}{2^7\cdot3}+
\frac{11x^6}{2^4\cdot3\cdot5}+\frac{841x^7}{2^9\cdot3^2\cdot5}+
\frac{151x^8}{2^4\cdot3^2\cdot5\cdot7}+\cdots. \eqno(3.20)
$$

Differentiating (3.19) we have
$$
\sum^{\infty}_{m=1}J'_m(mx)=\frac{1}{2\pi}\int_0^{\pi}d\theta\cot\left(\frac{\theta}{2}-
\frac{x}{2}\sin\theta\right)\sin\theta.                \eqno(3.21)
$$
Differentiating (3.20) we find
$$
\sum^{\infty}_{m=1}J'_m(mx)=\frac12+\frac{x}{2}+\frac{3x^2}{2^3}
+\frac{x^3}{3}+\frac{23\cdot5x^4}{2^7\cdot3}+
\frac{11x^5}{2^3\cdot5}+\frac{841\cdot7x^6}{2^9\cdot3^2\cdot5}+
\frac{151x^7}{2\cdot3^2\cdot5\cdot7}+\cdots. \eqno(3.22)
$$

For $1-x^2<<1, m>>1$ we have, (see (9.3.43) and (10.4.16) in
Abramowitz and Stegun, 1964)
$$
J'_m(mx)=-\left(\frac{2}{m}\right)^{2/3}Ai'\left(\left(\frac{m}{2}\right)^{2/3}(1-x^2)\right)=
\frac{1-x^2}{\sqrt3\pi}K_{2/3}\left(\frac{m}{3}(1-x^2)^{3/2}\right).
                                                     \eqno(3.23)
$$
Using this in the l.h.s. of (3.21) and replacing the summation by
integration we get
$$
\sum^{\infty}_{m=1}J'_m(mx)=2\sum^{\infty}_{n=1}J'_{2n}(2nx)=\frac{\sqrt3}{(1-x^2)^{1/2}}, \quad
1-x^2<<1,\quad m>>1.                            \eqno(3.24)
$$
Here we used the relation (see (651.16) in Gradsteyn and Ryzhic, 2000)
$$
\int_0^{\infty}dxx^{\alpha-1}K_{\nu}(ax)=2^{\alpha
-2}a^{-\alpha}\Gamma\left(\frac{\alpha+\nu}{2}\right)\Gamma\left(\frac{\alpha-\nu}{2}\right).
                                                            \eqno(3.25)
$$

Differentiating (3.24) we find
$$
\sum^{\infty}_{m=1}mJ''_m(mx)=\frac{\sqrt3}{(1-x^2)^{3/2}}, \quad
1-x^2<<1,\quad m>>1.                            \eqno(3.24')
$$
 Now we want to obtain the correction terms to (3.24)
explicitly. To this end we rewrite the integral in the r.h.s. of
(3.21) as follows
$$
\int_0^{\pi}d\theta\cot\varphi(\theta,x)\sin\theta=\int_0^{\pi}d\theta
[\frac{1}{\varphi}+(\cot\varphi-\frac{1}{\varphi})]\sin\theta,\quad
\varphi=\frac{\theta}{2}-\frac{x}{2}\sin\theta.
                                                    \eqno(3.26)
$$
In the same manner
$$
\int_0^{\pi}d\theta\frac{\sin\theta}{\varphi}=\frac2x\int_0^{\pi}d\theta
\frac{(x\sin\theta-\theta)+\theta}{\theta-x\sin\theta}=
\frac{2}{x}[-\pi+\int_0^{\pi}d\theta\frac{\theta}{\theta-x\sin\theta}].
                                                          \eqno(3.27)
$$
Taking into account that
$$
\theta-x\sin\theta=\frac{x\theta}{3!}[c+\theta^2-
\frac{3!\theta^4}{5!}+\frac{3!\theta^6}{7!}-\cdots],
   \quad c=6\frac{1-x}{x}                                                \eqno(3.28)
$$
we may write
$$
\int_0^{\pi}d\theta\frac{\theta}{\theta-x\sin\theta}=
\int_0^{\pi}d\theta[\left(\frac{\theta}{\theta-x\sin\theta}-\frac{6}{x(c+\theta^2)}
\right)+\frac{6}{x(c+\theta^2)}].                \eqno(3.29)
$$
The last term on the r.h.s. can be evaluated with the help of
formula
$$
I_1=\int_0^{\pi}d\theta\frac{1}{c+\theta^2}=\frac{1}{\sqrt
c}\arctan\frac{\pi}{\sqrt c},                       \eqno(3.30)
$$
see (120.01) in Dwight, 1961. Finally we have the exact relation
$$
\sum^{\infty}_{m=1}J'_m(mx)=-\frac1x+\frac{1}{2\pi}
\int_0^{\pi}d\theta[\sin\theta\left(cot\varphi-\frac{1}{\varphi}\right)
+\frac2x\left(\frac
{\theta}{\theta-x\sin\theta}-\frac{6}{x(c+\theta^2)}\right)]+
$$
$$
\frac{6}{\pi x^2\sqrt c}\arctan\frac{\pi}{\sqrt c},\quad
\varphi=\frac{\theta}{2}-\frac{x}{2}\sin\theta.      \eqno(3.31)
$$
 The integral here is of order unity and can be tabulated. For
 $1-x^2<<1$ the leading term in the r.h.s. is
 $$
\frac{6}{\pi \sqrt c}\arctan\frac{\pi}{\sqrt c}\approx
\frac{6}{\pi \sqrt c}\frac{\pi}{2}=\frac{3}{\sqrt
c}\approx\sqrt{\frac{3}{1-x^2}}                    \eqno(3.32)
 $$
as it should be according to (3.24).

Similarly to (3.31) we find
$$
\sum^{\infty}_{n=1}J'_{2n}(2nx)=-\frac{1}{2x}+\frac{1}{2\pi}
\int_0^{\pi}d\theta[\sin\theta\left(cot\psi-\frac{1}{\psi}\right)+\frac1x\left(\frac
{\theta}{\theta-x\sin\theta}-\frac{6}{x(c+\theta^2)}\right)]+
$$
$$
\frac{3}{\pi x^2\sqrt c}\arctan\frac{\pi}{\sqrt c},\quad
\psi=\theta-x\sin\theta.                              \eqno(3.31')
$$

 Now, if we differentiate (3.21), we get
$$
\sum^{\infty}_{m=1}mJ''_m(mx)=\frac{1}{4\pi}\int_0^{\pi}d\theta\left(
\frac{\sin\theta}{\sin\varphi}\right)^2.           \eqno(3.33)
$$
On the other hand, using Bessel equation
$$
mJ''_m(mx)=\left(\frac{1}{x^2}-1\right)mJ_m(mx)-\frac1xJ'_m(mx),\eqno(3.34)
$$
we have
$$
\sum^{\infty}_{m=1}mJ''_m(mx)=\left(\frac{1}{x^2}-1\right)\sum^{\infty}_{m=1}mJ_m(mx)
-\frac1x\sum^{\infty}_{m=1}J'_m(mx),
                                                        \eqno(3.35)
$$
This can be written in the form 
$$
\left(\frac{d}{dx}+\frac1x\right)\sum^{\infty}_{m=1}mJ'_(mx)=
\left(\frac{1}{x^2}-1\right)\sum^{\infty}_{m=1}mJ(mx). \eqno(3.35')
$$

It is desirable to present the the l.h.s. of this equation  in the
form similar to the r.h.s of (3.31). It turned out rather cumbersome
and we only indicate how to do this. As before we repeatedly use
"the subtract and add rule". Namely, to improve the behavior of a
term we subtract its asymptotic and add it. The term with subtracted
asymptotic we write usually in round brackets. The added term as a
rule can be integrated explicitly.
So we write
$$
\frac{1}{\sin\varphi}=\csc\varphi= \frac{1}{\varphi}+\frac16\varphi+\frac{7}{360}\varphi^3+\cdots,
$$
i.e. $\frac{1}{\sin\varphi}-\frac{1}{\varphi}$ has good behavior for
$\varphi\to 0$. so we may write
$$
\frac{\sin^2\theta}{\sin^2\varphi}=\sin^2\theta[\left(
\frac{1}{\sin\varphi}-\frac{1}{\varphi}
\right)+\frac{1}{\varphi}]^2=
$$
$$
\sin^2\theta[\left( \frac{1}{\sin\varphi}-\frac{1}{\varphi}
\right)^2+\frac{2}{\varphi}\left(
\frac{1}{\sin\varphi}-\frac{1}{\varphi}\right)+\frac{1}{\varphi^2}].
                                                        \eqno(3.36)
$$
As in (3.27) we write
$$
\frac{\sin\theta}{\varphi}=
\frac{2}{x}\left(-1+\frac{\theta}{\theta-x\sin\theta}\right).
\eqno(3.37)
$$
Proceeding further in this way we arrive at the expression
$$
\sum^{\infty}_{m=1}mJ''_m(mx)=\frac{1}{4\pi}\int_0^{\pi}d\theta\{
\left(\frac{1}{\sin\varphi}-\frac{1}{\varphi}\right)^2\sin^2\theta
+\frac{2\sin^2\theta}{\varphi}\left(\frac{1}{\sin\varphi}-\frac{1}{\varphi}\right)+
$$
$$
+\frac{4}{x^2}\left((\frac
{\theta}{\theta-x\sin\theta}-\frac{6}{x(c+\theta^2)}
\right)^2+\frac{4}{x^2}[1-2\left((\frac
{\theta}{\theta-x\sin\theta}-\frac{6}{x(c+\theta^2)}
\right)+\frac{6}{x(c+\theta^2)}]
$$
$$
+\frac{48}{x^3(c+\theta^2)}[\left(\frac
{\theta}{\theta-x\sin\theta}-\frac{6}{x(c+\theta^2)}
 -\frac{6^2\theta^4}{5!x(c+\theta^2)^2}\right)+\frac{6^2\theta^4}{5!x(c+\theta^2)^2}]+
%$$
%$$
 \frac{2^4\cdot3^3}{x^4(c+\theta^2)^2}\}.   \eqno(3.38)
$$
The term in the last squire bracket   is twice improved because it
is multiplied by large factor $\frac{48}{x^3(c+\theta^2)}$

As said above the integrals with compensating terms in the integrand can be evaluated
explicitly. The leading term in (3.38) (for $c<<1$) contain integral
$$
 I_2= \int_0^{\pi}d\theta\frac{1}{(c+\theta)^2}=\frac{1}{2c^{3/2}}\arctan\frac{\pi}{\sqrt
 c}+\frac{\pi}{2c(\pi^2+c)}.   \eqno(3.39)
$$
So the largest term in the r.h.s. of (3.38) is
$$
\frac{1}{4\pi}\frac{2^4\cdot3^2}{x^4}I_2. \eqno(3.40)
$$
For $c<<1$ this agrees with (3.24'). The integral (3.39) can be
obtained by differentiating (3.30) over $c$.

The next to the largest contributions to (3.38) contain integral
(3.30) and
$$
I^{(4)}_3=\int_0^{\pi}d\theta\frac{\theta^4}{(c+\theta^2)^3}=\int_0^{\pi}d\theta
\frac{1}{(c+\theta^2)^3}[(\theta^2+c)-c]^2. \eqno(3.41)
$$
To evaluate this integral we need (3.30), (3.39) and
$$
I_3=\int_0^{\pi}d\theta\frac{1}{(c+\theta^2)^3}=
 \frac{3}{8c^{5/2}}\arctan\frac{\pi}{\sqrt
 c}+\frac{3\pi}{8c^2(\pi^2+c)}+\frac{\pi}{4c(\pi^2+c)^2}, \eqno(3.42)
$$
which can be obtained by differentiating $I_2$. Using these
integrals we find
$$
I_3^{(4)}=\int_0^{\pi}d\theta\frac{\theta^4}{(c+\theta^2)^3}=
\frac{3}{8\sqrt c}\arctan\frac{\pi}{\sqrt
c}-\frac{5\pi}{8(\pi^2+c)}+\frac{c\pi}{4(\pi^2+c)}.\eqno(3.43)
$$
The contribution to (3.38) from this term is
$$
\frac{1}{4\pi}\frac{2^6\cdot3^3}{5!x^4}I_3^{(4)}. \eqno(3.44)
$$
The remaining terms in (3.38) should give integral of order unity.
The sum $\sum^{\infty}_{m=1}mJ_m(mx)$ can be obtained from (3.35),
(3.38), and (3.31).

We note that integral $I_3^{(4)}$ is a special case of (1.2.10.5) in
Prudnikov et al., 1986, but it seems there is something wrong with
that formula.

The case $m=2n$ is treated similarly to just considered one:
$$
\sum^{\infty}_{n=1}\frac{1}{2n}J_{2n}(2nx)=-\frac{1}{2\pi}\int_0^{\pi}d\theta
\ln(2\sin\psi), \quad \psi=\theta-x\sin\theta.
                                                      \eqno(3.45)
$$
Differentiating we get
$$
\sum^{\infty}_{n=1}J'_{2n}(2nx)=\frac{1}{2\pi}\int_0^{\pi}d\theta
\cot\psi\sin\theta, \quad \psi=\theta-x\sin\theta.
                                                      \eqno(3.46)
$$
Differentiating once more we obtain
$$
\sum^{\infty}_{n=1}2nJ''_{2n}(2nx)=\frac{1}{2\pi}\int_0^{\pi}d\theta
\frac{\sin^2\theta}{\sin^2\psi}.
                                                      \eqno(3.47)
$$
Now
$$
\frac{\sin^2\theta}{\sin^2\psi}=\left(\frac{1}{\sin\psi}-
\frac{1}{\psi}\right)^2\sin^2\theta
+\frac{2\sin^2\theta}{\psi}\left(\frac{1}{\sin\psi}-\frac{1}{\psi}\right)+
$$
$$
+\frac{1}{x^2}\langle1-2[\left(\frac
{\theta}{\theta-x\sin\theta}-\frac{6}{x(c+\theta^2}
\right)+\frac{6}{x(c+\theta^2)}]+ \left(\frac
{\theta}{\theta-x\sin\theta}-\frac{6}{x(c+\theta^2)}
\right)^2+
$$
$$
\frac{12}{x(c+\theta^2)}[
\left(\frac{\theta}{\theta-x\sin\theta}-\frac{6}{x(c+\theta^2)}
 -\frac{3\theta^4}{10x(c+\theta^2)^2}\right)+\frac{3\theta^4}{10x(c+\theta^2)^2}]+
%$$
%$$
 \left(\frac{6}{x(c+\theta^2)}\right)^2\rangle.   \eqno(3.48)
$$
Not only the leading term is half of that in (3.38) but the same is
true for the subleading terms. which slightly decrease the leading
term.

For small $x$ we have the expansions in power series of $x$. It is easy to find
$$
\sum^{\infty}_{n=1}2nJ_{2n}(2nx)=x^2+\frac{7x^4}{3}+
\frac{239x^6}{2^2\cdot3\cdot5}+\frac{1481x^8}{2^2\cdot3^2\cdot7}+
\frac{292223x^{10}}{2^6\cdot3^4\cdot7}+\cdots. \eqno(3.49)
$$
For larger $x$ see (3.55) below.

Retaining only terms with even powers of $x$ in
(3.20) we get
$$
\sum^{\infty}_{n=1}\frac{1}{2n}J_{2n}(2nx)=\frac{x^2}{2^2}+\frac{x^4}{2^2\cdot3}
+ \frac{11x^6}{2^4\cdot3\cdot5}+
 \frac{151x^8}{2^4\cdot3^2\cdot5\cdot7}+\frac{15619x^{10}}{2^8\cdot3^4\cdot5\cdot7}
 +\cdots. \eqno(3.50)
$$
Differentiating (3.50) we have
$$
\sum^{\infty}_{n=1}J'_{2n}(2nx)=\frac{x}{2}+\frac{x^3}{3}+
\frac{11x^5}{2^3\cdot5}+\frac{151x^7}{2\cdot3^2\cdot5\cdot7}
+\frac{15619x^9}{2^7\cdot3^4\cdot7}+\cdots     \eqno(3.51)
$$

Taking into account (3.24) we may rewrite (3.51) in the form
$$
\sum^{\infty}_{n=1}J'_{2n}(2nx)=\frac{x}{2\sqrt{1-x^2}}[1+\frac{x^2}{2\cdot3}+
\frac{11x^4}{2^3\cdot3\cdot5}+\frac{59x^6}{2^4\cdot3^2\cdot7}+
\frac{14971x^8}{2^7\cdot3^4\cdot5\cdot7}+\cdots]. \eqno(3.52)
$$

In Section 6 we also need
$$
\int_0^{\beta}dx\sum^{\infty}_{n=1}nJ_{2n}(2nx)=\frac{1}{4\sqrt3(1-\beta^2)^{3/2}}, \quad 1-\beta<<1.\eqno(3.53)
$$ 
Differentiating this over $\beta$ we get
$$
\sum^{\infty}_{n=1}nJ_{2n}(2n\beta)=\frac{\sqrt3}{4(1-\beta^2)^{5/2}}, \eqno(3.54)
$$
which can be verified similarly to (3.24), using (5.03) below.
 We  also find
$$
\sum^{\infty}_{n=1}2nJ_{2n}(2nx)=\frac{x^2}{(1-x^2)^{5/2}}[1-\frac{x^2}{2\cdot3}+
\frac{x^4}{2^3\cdot5}+\frac{5x^6}{2^4\cdot3^2\cdot7}+
\frac{103x^8}{2^7\cdot3^4\cdot7}+\cdots]. \eqno(3.55)
$$
and 
$$
\int_0^{\beta}dx\sum^{\infty}_{n=1}nJ_{2n}(2nx)=
\frac{\beta^3}{6(1-\beta^2)^{3/2}}[1-\frac{\beta^2}{2\cdot5}-
\frac{\beta^4}{2^3\cdot7}-\frac{19\beta^6}{2^4\cdot3^3\cdot7}-
\frac{809\beta^8}{2^7\cdot3^3\cdot7\cdot11}-\cdots ]. \eqno(3.56)
$$

\section{Bilinear algebraic series}

Most of the results in this case are contained in Watson. 1966. see
also Peters and Mathews, 1963 and Lerche and Tautz, 2007. We note
here two misprints in the last but one paper. The factor in front of
$\Sigma\Sigma$ on page 439 (in equation that should be (A2)) should
be $(1-e^2)^2$ not $(1-e^2)$. In equation (A1) the factor just after
$\frac{J'^2_n}{n^2}$ should be $\left(\frac{4}{e}\right)^2$ not
$\left(\frac{4}{e^2}\right)^2$.

We start with easily verifiable relation, see Schott, 1912.
$$
(1-x^2)\frac{d}{dx}J^2_n(nx)=\frac{d}{dx}[x^2J'^2_n(nx)].\eqno(4.01)
$$
To check it we simply perform differentiation and use Bessel
equation. Multiplying (4.01) by $n^{\nu}$ and summing over $n$, we
have
$$
(1-x^2)\frac{d}{dx}\Sigma_1^{\infty}n^{\nu}J^2_n(nx)=
\frac{d}{dx}[x^2\Sigma_1^{\infty}n^{\nu}J'^2_n(nx)]. \eqno(4.02)
$$
Putting $\nu=-2$ we get
$$
(1-x^2)\frac{d}{dx}\Sigma_1^{\infty}\frac{1}{n^2}J^2_n(nx)=
\frac{d}{dx}[x^2\Sigma_1^{\infty}\frac{1}{n^2}J'^2_n(nx)].
                                                          \eqno(4.03)
$$
Now we can use in the l.h.s. the Nielson formula
$$
\Sigma_1^{\infty}\frac{1}{n^2}J^2_n(nx)=\frac{x^2}{4}. \eqno(4.04)
$$
Performing differentiation in the l.h.s. of (4.03) we obtain
$$
(1-x^2)\frac
x2=\frac{d}{dx}[x^2\Sigma_1^{\infty}\frac{1}{n^2}J'^2_n(nx)].
                                                               \eqno(4.05)
$$
Integrating over $x$ we find the analog of Nielson formula
$$
\Sigma_1^{\infty}\frac{1}{n^2}J'^2_n(nx)=\frac14-\frac{x^2}{8}.
                                                            \eqno(4.06)
$$

Sums of any two algebraic series are related algebraically. For
example (see (7.15 48) in Bateman and Erdely, 1953)
$$
\Sigma_1^{\infty}J^2_n(nx)=\frac12[\frac{1}{\sqrt{1-x^2}}-1]\eqno(4.07)
$$
and
$$
\Sigma_1^{\infty}J'^2_n(nx)=\frac{1}{2x^2}[1-\sqrt{1-x^2}]
                                                             \eqno(4.08)
$$
are related as follows
$$
\frac{\sqrt{1-x^2}}{x^2}\Sigma_1^{\infty}J^2_n(nx)=\Sigma_1^{\infty}J'^2_n(nx).
                                                                \eqno(4.09)
$$
Equations (4.07) and (4.08) are easily obtainable by the method of Peters and Mathews, 1963. 
Differentiating (4.07) we obtain
$$
\Sigma_1^{\infty}nJ'_n(nx)J_n(nx)=\frac{x}{4(1-x^2)^{3/2}}.
                                                       \eqno(4.10)
$$

Differentiating this equation we have
$$
\Sigma_1^{\infty}[n^2J'^2_n(nx)+n^2J_n(nx)J''_n(nx)]=\frac{1+2x^2}{4(1-x^2)^{5/2}}.
                                                             \eqno(4.11)
$$
Using here Bessel equation
$$
nJ''_n(nx)=\left(\frac{1}{x^2}-1\right)nJ_n(nx)-\frac1xJ'_n(nx)
                                                            \eqno(4.12)
$$
and (4.10) we find
$$
\Sigma_1^{\infty}[n^2J'^2_n(nx)+\left(\frac{1}{x^2}-1\right)n^2J^2_n(nx)]=
\frac{2+x^2}{4(1-x^2)^{5/2}}.
                                                     \eqno(4.13)
$$

Next, differentiating the relation, see Schott, (1912)
$$
\Sigma_1^{\infty}n^2J^2_n(nx)=\frac{4x^2+x^4}{16(1-x^2)^{7/2}}
                                                         \eqno(4.14)
$$
we have
$$
\Sigma_1^{\infty}2n^3J_n(nx)J'_n(nx)=\frac
x2\frac{1+3x^2+\frac38x^4}{(1-x^2)^{9/2}}.
                                                        \eqno(4.15)
$$
Differentiating it, using Bessel equation (4.12), and 
equation (4.15) we find
$$
\Sigma_1^{\infty}[n^4J'^2_n(nx)+\left(\frac{1}{x^2}-1\right)n^4J^2_n(nx)]=
\frac{1}{(1-x^2)^{11/2}}[\frac{1}{2}+\frac{19}{4}x^2+\frac{69}{16}x^4+\frac{9}{32}x^6].
                                                     \eqno(4.16)
$$

To obtain more general relation we twice  differentiate $J^2_n(y)$:
$$
\frac{d}{dy}J_n^2(y)=2J_n(y)J'_n(y),\quad
\frac{d^2}{dy^2}J_n^2(y)=2{J'}^2_n(y)+2J_n(y)J''_n(y)=
$$
$$
2{J'}^2_n(y)+2J_n(y)[\left(
 \frac{n^2}{y}-1\right)J_n(y)-\frac1yJ'_n(y)].
$$
From here, see eq. (40.4) in Iwanenko and Sokolov (1951)
$$
{J'}^2_n(y)=[\frac12\left(\frac1y\frac{d}{dy}+\frac{d^2}{dy^2}
\right)+\left(1-\frac{n^2}{y^2}\right)]J_n^2(y).       \eqno(4.17)
$$
Putting $y=nx$, multiplying by $n^{\nu}$ and summing over $n$ we get

$$
\Sigma_1^{\infty}n^{\nu}{J'}^2_n(y)=\frac12\left(\frac1x\frac{d}{dx}+\frac{d^2}{dx^2}
\right)\Sigma_1^{\infty}n^{\nu-2}J_n^2(nx)+
\left(1-\frac{1}{x^2}\right)\Sigma_1^{\infty}n^{\nu}J_n^2(nx).
                                                          \eqno(4.18)
$$
More generally we could multiply by $a_n$ instead of $n^{\nu}$.

\section{Bilinear transcendental series}

To deal with power series expansions we use the equation (8.442.1) in
Gradshteyn and Ryzhik, 2000:
$$
J_n^2(y)=\Sigma_1^{\infty}(-1)^s\frac{(2n+2s)!y^{2(n+s)}}
{s!2^{2(n+s)}(2n+s)![(n+s)!]^2}.
                                                \eqno(5.01)
$$
Similar expansion for ${J'}^2_n(nx)$ can be obtained from (4.17) but
it is simpler to use (4.01) or directly (4.02) if we need a sum. Using (5.01) we can
find
$$
\Sigma_1^{\infty}nJ_n^2(nx)=\frac{x^2}{4}\left(1+\frac{7x^2}{2^2}+
 \frac{239x^4}{2^5\cdot3}+ \frac{7435x^6}{2^8\cdot3^2}+
 \frac{292223x^8}{2^{13}\cdot3^2}+\cdots\right).      \eqno(5.02)
$$
Much easier to obtain it  from (3.49) using (2.01). For $1-x^2<<1$ we
use the relation, see  (9.3.35) and (10.4.14) in Abramowitz and Stegun, 1964
$$
J_n(nx)=\frac{\sqrt{1-x^2}}{\pi\sqrt3}K_{1/3}[\frac
n3(1-x^2)^{3/2}],                                  \eqno(5.03),
$$
 replace sum by integral and use (6.576 4) in Gradshteyn and Ryzhic to
 get
 $$
\left.\Sigma_1^{\infty}nJ_n^2(nx)\right|_{1-x^2<<1}=
\frac{1}{\pi\sqrt3}\frac{1}{(1-x^2)^2}.
                                                 \eqno(5.04)
 $$
This suggest the form
$$
\Sigma_1^{\infty}nJ_n^2(nx)=\frac{1}{(1-x^2)^2}\Sigma_1^{\infty}c_{2n}x^{2n}.\eqno(5.05)
$$
We chose $c_{2n}$ so as to agree with (5.02). Then we get
$$
\Sigma_1^{\infty}nJ_n^2(nx)=\frac{x^2}{4(1-x^2)^2}[1-\frac{x^2}{2^2}-
\frac{x^4}{2^5\cdot3}
-\frac{5x^6}{2^8\cdot3^2}--\frac{23x^8}{2^{13}\cdot3^2}-\cdots].
\eqno(5.06)
$$

Proceeding in the same manner we find
$$
\Sigma_1^{\infty}n{J'}^2_n(nx)=\frac{1}{2^2}+
\frac{5x^2}{2^4}+\frac{127x^4}{2^7\cdot3}+
\frac{3133x^6}{2^{10}\cdot3^2}+\frac{101887x^8}{2^{15}\cdot3^2}+\cdots.\eqno(5.07)
$$
Using (3.23) with $m=2n$ and equation (6.576 4) in Gradshteyn and Ryzhic we get
$$
\left.\Sigma_1^{\infty}n{J'}^2_n(nx)\right|_{1-x^2<<1}=
\frac{2}{\pi\sqrt3}\frac{1}{1-x^2}.
                                                 \eqno(5.08)
$$

Similarly to (5.06) we find
$$
\Sigma_1^{\infty}n{J'}^2_n(nx)=\frac{1}{4(1-x^2)}[1+\frac{x^2}{2^2}+
+\frac{7x^4}{2^5\cdot3}+\frac{85x^6}{2^8\cdot3^2}+\frac{1631x^8}{2^{13}\cdot3^2}+\cdots].
                                                          \eqno(5.09)
$$

With the help of (5.01) we can find that
$$
\Sigma_1^{\infty}\frac1nJ_n^2(nx)=\frac{x^2}{4}[1+\frac{x^2}{2^2}+
 \frac{11x^4}{2^5\cdot3}+\frac{151x^6}{2^8\cdot3^2}+
 \frac{15619x^8}{2^{13}\cdot3^2\cdot5}+\cdots].       \eqno(5.10)
$$
Much easier to get it using (3.50) and (2.01). With the help of
(2.01) and (3.19) we obtain
$$
\Sigma_1^{\infty}\frac1nJ_n^2(nx)=-\frac{1}{\pi^2}\int_0^{\pi}d\psi\int_0^{\pi}d\theta
\ln[2\sin(\theta-x\sin\psi\sin\theta)]. \eqno(5.11)
$$
Differentiating this we get
$$
\Sigma_1^{\infty}2J_n(nx)J'_n(nx)=
\frac{1}{\pi^2}\int_0^{\pi}d\psi\int_0^{\pi}d\theta
\cot(\theta-x\sin\psi\sin\theta)\sin\psi\sin\theta. \eqno(5.12)
$$
\section{Application}

In classical case an electron in constant magnetic field or in
circularly polarized plain wave field moves (in proper coordinate
system) on circular orbit. The intensity of radiation is given by
Schott formula. see Landau and Lifshitz \S  74
$$
dI_n=\frac{e^2}{2\pi c}(n\omega_H)^2[\tan^2\theta
J_n^2(n\beta\cos\theta)+\beta^2 {J'}_n^2(n\beta\cos\theta)]d\Omega, \quad \beta=\frac {v}{c}.
\eqno(6.01)
$$
Integrated over angles intensity is
$$
I_n=\frac{2e^2}{v}\omega_H^2[\beta^2nJ'_{2n}(2n\beta)-
(1-\beta^2)\int_0^{\beta}dxn^2J_{2n}(2nx)].\eqno(6.02)
$$
In case of magnetic field
$\omega_H=\frac{eH}{mc}\sqrt{1-\beta^2}$.

 The probability (per unit time) is obtained by dividing the intensity by $
\hbar n\omega_H$, $\omega_H$ is the frequency of the first harmonic:
$dP_n=dI_n/\hbar n\omega$. 
So, (6.01) divided by $\hbar n\omega_H$ and summed up over $n$ leads us to series (5.06) and (5.09).

For total pobability (per unit time) we
find
$$
P=\Sigma_1^{\infty}P_n=\frac{2e^2}{\hbar
v}\omega_H[\beta^2\Sigma_1^{\infty}J'_{2n}(2n\beta)-
(1-\beta^2)\int_0^{\beta}dx\Sigma_1^{\infty}nJ_{2n}(2nx)].\eqno(6.03)
$$
We see that summing over $n$ the expressions for intensity gives
algebraic series, while summing probabilities leads to trancendental series.

 The probability
that no radiation occurs during time $t$ is $\exp(-Pt)$. Using
(3.52) and (3.56) we find
$$
P=\frac{2e^2}{\hbar
 v}\frac{\omega_H}{\sqrt{1-\beta^2}}\frac{\beta^3}{3}[1+\frac{3\beta^2}{2\cdot5}+
\frac{41\beta^4}{2^3\cdot5\cdot7}+
\frac{5^2\cdot11\beta^6}{2^4\cdot3^3\cdot7}+
\frac{28121\beta^8}{2^7\cdot3^2\cdot5\cdot7\cdot11}+\cdots].
                                                           \eqno(6.04)
$$
In case of circularly polarized wave $\omega_H$ is the frecuency of wave. 
 For $1-\beta<<1$ we find
$$
P=\frac{e^2}{\hbar c}\frac{5}{2\sqrt3}\frac{eH}{mc},\quad
\frac{e^2}{\hbar c}=\alpha=1/137.
                                                \eqno(6.05)
$$
So for ultrarelativistic electron $P$ in classical case is
independent of energy. But it is known that in this case the
radiation is the same in any external field, slowly verying on the formation length of radiation [see equations (27')
and (27'') in  Nikishov
and Ritus, 1964] and is given by
$$
 W=\frac{5\alpha m}{2\sqrt3}\chi, \quad
 \chi=\frac{\sqrt{(eF_{\mu\nu}p^{\nu})^2}}{m^3}. \eqno(6.06)
$$
Here $W$ is the probability per unit proper time
$\tau=\sqrt{1=\beta^2}t$,  $F_{\mu\nu}$ is the field tensor and $p$
electron momentum, $\hbar =c=1$, $Pt=W\tau$. For the magnetic field
$\chi=\frac{eHv}{m^2\sqrt{1-\beta^2}}$ and we see that (6.05) is in
agreement with (6.06).

It is interesting to note that in quantum case. when $\chi>>1$
$$
W=\frac{14\Gamma(2/3)\alpha m}{27}(3\chi)^{2/3}. \eqno(6.07)
$$
This means the free from radiation (laboratory) lifetime increases
as $\left(\frac{E}{m}\right)^{1/3}$.
\section*{Conclusion}
It seems that much more can be done in the considered region by
mathematicians.
\section*{References}

Abramowitz M. and Stegun I., {\sl Handbook of Mathematical Functions},
National Bureau of Standards, 1964.\\
 Bateman H., and Erdelyi A. {\sl Higher Transcendental Functions}, Vol.
 2, New York, Mc Grow-Hill, 1954.\\
 Dwight  H. B.{\sl Tables of integrals and other
mathematical data}, 1961.\\
New York, the Macmillan company, 1961.\\
Gradshteyn I.S., and Ryzhic I.N.,{\sl Tables of Integrals, Series and
Products}, London, Academic, 2000.\\
Ivanenko D.D., and Sokolov A.A., {\sl Classical Theory of Fields}. (in
Russian), Moscow, 1951.\\
Landau L.D., and  Lifshitz, {\sl Classical Theory of Fields}, Oxford:
Butterworth-Heineman, 1994.\\
Lerche I., and Tautz R.C., Astrophys. Jour.,665, p.1288, 2007.\\
Nikishov A.I. and Ritus V.I.,JETP, Vol 46, p. 776, 1964\\
Peters P.C., and  Mathews J., Phys. Rev.,{\bf 131}, 435, 1963.\\
Prudnikov A.P., Brychkov Y.A., and Marichev O.I. {\sl Integrals and
Series},Vol.I, Gordon and Breach, New York, 1986.\\
Schott G.A. {\sl Electromagnetic radiation}, Cambridge Univ. Press,
1912.\\
 Watson G.N., {\sl A Treaty on the Theory of Bessel Functions}:
Cambridge Univ. Press, 1966.\\
\end{document}